\documentclass{iau} 
\usepackage{graphicx}
\usepackage{natbib}
\usepackage{hyperref}
\usepackage{color, soul}

\title[Wind-ISM interaction of massive stars] 
{Interaction between massive star winds and the interstellar medium}

\author[Jonathan Mackey]   
{Jonathan Mackey$^{1}$}

\affiliation{$^1$Dublin Institute for Advanced Studies, Astronomy \& Astrophysics Section, DIAS Dunsink Observatory, Dublin, D15 XR2R, Ireland}

\pubyear{2022}
\volume{370}  
\jname{IAU Symposium 370: Winds from stars and exoplanets}
\editors{ADD EDITORS}
\begin{document}

\maketitle

\begin{abstract}
Massive stars drive strong winds that impact the surrounding interstellar medium, producing parsec-scale bubbles for isolated stars and superbubbles around young clusters.
These bubbles can be observed across the electromagnetic spectrum, both the wind itself and the swept up interstellar gas.
Runaway massive stars produce bow shocks that strongly compresses interstellar gas, producing bright infrared, optical and radio nebulae.
With the detection of non-thermal radio emission from bow shocks, particle acceleration can now also be investigated.
I review research on wind bubbles and bow shocks around massive stars, highlighting recent advances in infrared, radio and X-ray observations, and progress in multidimensional simulations of these nebulae.
These advances enable quantitative comparisons between theory and observations and allow to test the importance of some physical processes such as thermal conduction and Kelvin-Helmholtz instability in shaping nebulae and in constraining the energetics of stellar-wind feedback to the interstellar medium.
\end{abstract}

\firstsection 
\section{Models for spherical wind bubbles}
Massive stars have a strong effect on their surroundings through their intense radiation (especially extreme-UV, ionizing radiation), strong winds, eruptive explosions and supernova explosions at the end of their lives.
Quantifying these feedback effects is important for understanding the dynamical and chemical evolution of galaxies, and also the structure and dynamics of the interstellar medium (ISM) in our own Galaxy.
On smaller scales, modelling of circumstellar nebulae can also give clues as to the evolutionary history of some nearby massive stars.
Comparison of models with observations can give constraints on different physical processes in astrophysical plasmas, such as particle acceleration and thermal conduction.

In one of the first papers studying the effects of stellar winds on the ISM, \citet{Mat66} proposed that the optical cavity in the Rosette Nebula around NGC2244 is maintained by dynamical pressure of the strong and high-velocity winds of the massive stars in the central cluster.
\citet{DysDeV72} built on this work, developing a theory for the dynamics of wind-blown bubbles.
This was further generalised by \citet{CasMcCWea75}, and again in the classic paper by \citet{WeaMcCCas77}.
The basic picture is that of a spherical wind expanding from a star or group of stars and displacing the interstellar gas.
From the contact discontinuity between the two media, a reverse (or termination) shock propagates backwards towards the star generating a hot bubble of shocked coronal gas with temperature $T\sim10^6-10^8$\,K.
Similarly a forward shock propagates into the undisturbed ISM as long as the bubble expands supersonically.
The ISM is photoionized by extreme-UV radiation from the hot star(s), producing an H~\textsc{ii} region around the star that usually extends well beyond the wind bubble, but which may be trapped by the shocked ISM \citep{WeaMcCCas77, FreHenYor03}.

\citet{WeaMcCCas77} introduced thermal conduction, which smoothes out the contact discontinuity and produces significant quantities of gas in the temperature range from $10^5-10^7$\,K, that emits UV and thermal X-ray radiation.
This can reproduce the observed [O\,\textsc{vi}] line emission that is observed, but tends to overpredict thermal X-rays (see, e.g., \citealt{ToaOskGon16}).
The dynamical model of Weaver has the bubble radius, $r$, expanding with time, $t$, as $r\propto t^{3/5}$.
This is the solution in the adiabatic limit, i.e., in the limit that the thermal energy in the wind bubble is not lost and can power the pressure-driven expansion of the bubble.
In the momentum conserving limit, where the wind energy is efficiently radiated away, a simple dimensional analysis shows that $r\propto t^{1/2}$.
If thermal conduction is efficient at transporting thermal energy out of the bubble and into the mixing layer, then the second expansion law holds.

\citet{GarLanMac96} studied wind-ISM and wind-wind interactions with multi-dimensional simulations, showing that some shocks become effectively isothermal, forming thin and unstable layers that break up into clumps.
This multi-dimensional effect is important for the expansion rate and radiative emission of the nebulae, and can explain some of the observational properties of Wolf-Rayet nebulae.
This line of research was extended by \citet{FreHenYor06} with the inclusion of photoionizing radiative transfer, modelling the H~\textsc{ii} region and wind-bubble evolution for the full stellar lifetime, expanding into a uniform ISM.
They showed that dissipation and instabilities in expanding layers can significantly affect X-ray and UV emission from wind bubbles, potentially resolving the disagreement between large predicted X-ray luminosities, and relatively weak X-ray emission from observed nebulae.
\citet{ToaArt11} used different stellar evolution sequences and higher-resolution simulations, including calculations with and without thermal conduction, investigating the structure and X-ray emission from massive-star nebulae.
They concluded that both thermal conduction and dynamical mixing by hydrodynamic instabilities are affecting the X-ray emission.
\citet{GeeRosBla15} modelled the evolution of the CSM around a massive star including effects of both photoionization and stellar wind in 3D simulations with a similar setup.
The latest work using as initial conditions wind expanding from a star at rest with respect to a uniform ISM is by \citet{MeyPetPoh20}, who used high-resolution 2D simulations to study the CSM of a 60\,M$_\odot$ star through various evolutionary phases followed by explosion as a supernova.

\begin{figure}
\begin{center}
 \includegraphics[width=0.98\textwidth]{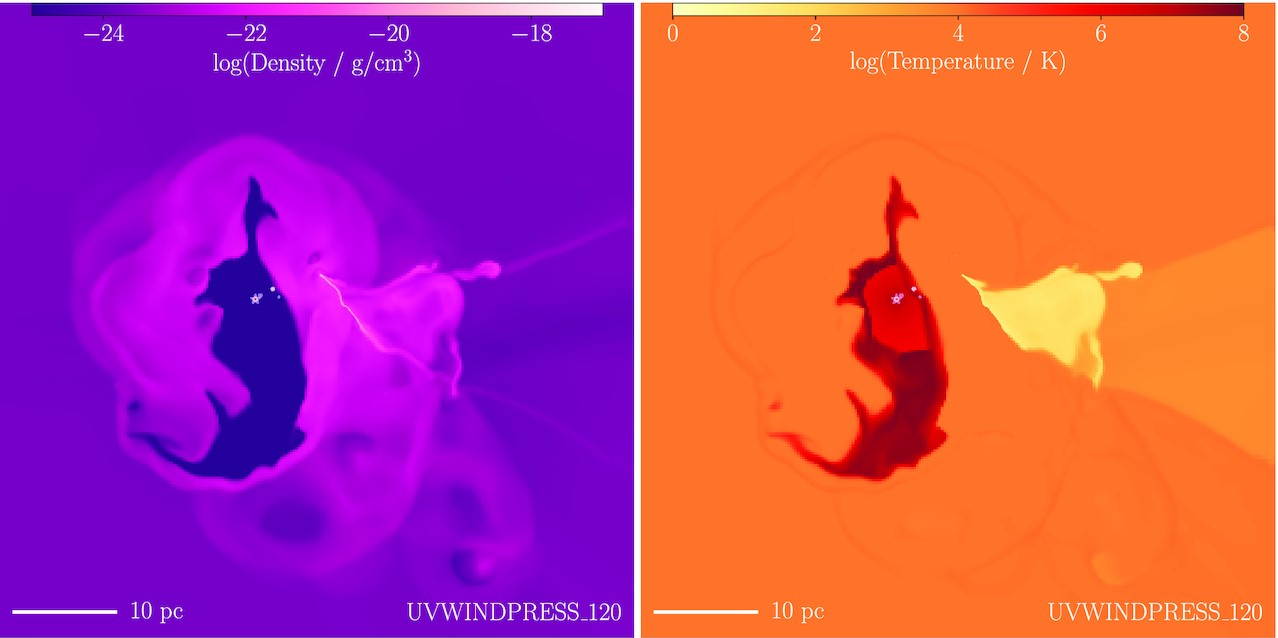} 
 \caption{Slice through a 3D simulation of a turbulent molecular cloud, 0.4\,Myr after a massive star has formed, showing gas density (left) and temperature (right).  The hot, low-density and asymmetric wind bubble surrounds the star.
 Reproduced from fig.~4 of ``The geometry and dynamical role of stellar wind bubbles in photoionized H~\textsc{ii} regions'', \citet{GeeBieRos21}, MNRAS, 501, 1352.}
   \label{fig:geen21}
\end{center}
\end{figure}

These simulations are very useful for elucidating the physical processes at work in wind bubbles, but they assume certain symmetries that are not always present.
In particular stellar motion or bulk flows in the ISM lead to distorted wind bubbles \citep{WeaMcCCas77, MacGvaMoh15} that produce a bow shock when the star is moving supersonically with respect to the ISM \citep[e.g.][]{VanMcC88}.
Furthermore the turbulent ISM is neither static nor homogeneous, and these density and velocity fluctuations also affect the shape of a wind bubble \citep[see Fig.~\ref{fig:geen21}, taken from][]{GeeBieRos21}.
The combination of stellar winds, radiative feedback, stellar clustering, and supernovae leads to a very complex multiphase ISM with strong pressure gradients on many scales \citep{RatNaaGir21}.

\section{Runaway Stars and bow shocks}
Not long after the first calculations showing that the interaction between Solar wind \citep{BarKraKul71} or stellar wind \citep{WeaMcCCas77} and local ISM could produce a bow shock, the bow shock around the closest O star to Earth, $\zeta$ Ophiuchi, was detected in nebular emission lines by \citet{GulSof79}.
Bow shocks provide an excellent laboratory for studying the wind-ISM interaction for a number of reasons:
\textit{(i)} the star has generally moved far from its place of birth and there are no other massive stars in the immediate vicinity that could induce wind-wind interactions;
\textit{(ii)} the star is typically moving in the diffuse (warm) phase of the ISM through a less structured medium than in a molecular cloud; and 
\textit{(iii)} ram pressure provided by stellar motion compresses the bow shock into a structure that is overdense (and more easily observable) and that has a relatively short dynamical timescale.

The structure of a bow shock is well-described in fig.~1 of \citet{ComKap98}, again consisting of two shocks separated by a contact discontinuity.
Stellar motion at velocity $v_\star$ with respect to the ISM leads to an asymmetric ISM ram pressure of $\rho_0 v_\star^2$ in the reference frame of the star ($\rho_0$ is the ISM gas density).
Pressure balance gives the characteristic size of the bow shock \citep{BarKraKul71}, the standoff distance, $R_0$, as
\begin{equation}
R_0 = \sqrt{\frac{\dot{M} v_\infty}{4\pi \rho_0 (v_\star^2 + v_\mathrm{A}^2 + c^2)}} \;,
\label{eqn:standoff}
\end{equation}
where $\dot{M}$ and $v_\infty$ are the mass-loss rate and terminal wind velocity of the star, respectively.
The Alfv\'en ($v_\mathrm{A}$) and sound ($c$) speeds represent the magnetic and thermal pressure contributions to the total pressure, but usually the ram pressure is the dominant term.
The dynamical timescale of the bow shock is $t_\mathrm{dyn} = R_0/v_\star$,  typically $\sim10^5$\,yr for bow shocks with $R_0\lesssim 1$\,pc and $v_\star\sim30$\,km\,s$^{-1}$, much shorter than the lifetime of a massive star.

This equation provides a measurement of $\dot{M}$ that is independent from stellar-atmosphere spectral-line studies \citep[e.g.][]{GulSof79}, if the other quantities on the right-hand side of the equation are well constrained.
\citet{GvaLanMac12} were able to use the properties of the H~\textsc{ii} region around the runaway O star $\zeta$ Oph to constrain $\rho_0$ and thereby measure $\dot{M}\approx2.2\times10^{-8}$\,M$_\odot$\,yr$^{-1}$, below the estimate derived from optical lines in the atmosphere and the \citet{VinDeKLam00} theoretical estimate, but well above the estimate derived from UV lines \citep{MarBouMar09}.
\citet{KobChiPov18} use Eq.~\ref{eqn:standoff} to measure $\dot{M}$ for a sample of bow shocks around 20 O stars by making the assumptions that all outer shocks are adiabatic with a density jump of $4\times$, that $\rho_0$ can be accurately measured from far-IR dust emission, and all stars are moving through the ISM with $v_\star=30$\,km\,s$^{-1}$.
This was followed up by \citet{KobChiPov19} who used \emph{Gaia} DR2 data to measure $v_\star$ for a larger sample of stars and thereby obtaining more accurate results.
\citet{HenArt19} made a careful analysis of the statistical and systematic uncertainties of measuring $\dot{M}$ from bow-shock observations, also comparing their method with that of \citet{KobChiPov18}.
They find that $\dot{M}$ measurements for individual sources may have large uncertainties (from uncertain dust properties and shocked ISM pressure support), but statistically for a large group of sources the methods are promising.

\subsection{Multiwavelength emission from bow shocks}
The first detection of a significant number of bow shocks was made with the \textit{IRAS} observatory by \citet{VanMcC88} in the mid-IR \citep[see also][]{VanNorDga95}.
The broadband IR emission is thermal radiation from interstellar dust grains photo-heated by the intense UV-radiation field of the massive star that drives the bow shock.
The \textit{AKARI} observatory obtained higher-resolution IR images of some bow shocks \citep{UetIzuYam08}, but a major advance came with the \textit{Spitzer} and \textit{WISE} missions.
The E-BOSS catalog of \citet{PerBenBro12, PerBenIse15} contains 73 confirmed and candidate bow shocks, and \citet{KobChiSch16} compiled a larger catalog of 709 candidate bow shocks selected by mid-IR morphology.
\citet{MeyMacLan14} showed that bow shocks are most luminous in the mid-IR \citep[see also][]{AcrSteHar16, HenArt19}, explaining why these surveys have been so successful.
Asymmetric wind bubbles with subsonic relative motion between star and ISM can also produce bright mid-IR arcs \citep{MacHawGva16}.

\begin{figure}
\begin{center}
 \includegraphics[width=1.0\textwidth]{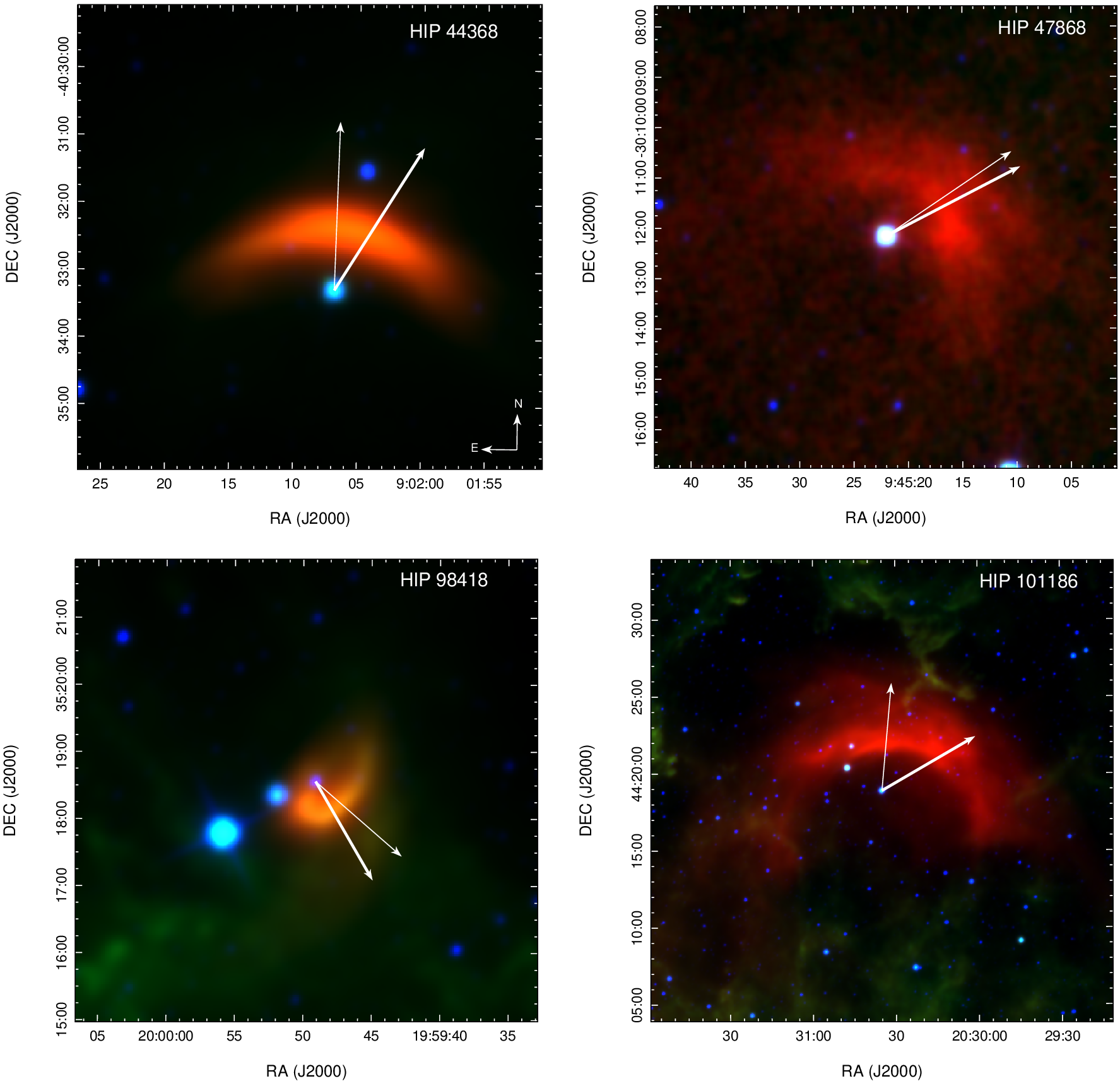} 
 \caption{Four bow shocks from the catalog of \citet{PerBenIse15}, imaged by \textit{WISE}. Red: band 4, 22.2\,$\mu$m. Green: band 3, 12.1\,$\mu$m. Blue: band 1, 3.4\,$\mu$m.
   Credit: Peri, Benaglia \& Isequilla, 2015, Astronomy \& Astrophysics, 578, A45, fig.~5.}
   \label{fig:peri15}
\end{center}
\end{figure}

Despite the first detection of a bow shock being in optical lines, this has proven to be a difficult method for detection because the massive stars are so optically bright that identifying nearby nebular emission is challenging \citep[see][]{GulSof79}.
A few other detections have been made: HD 165319 appears to show a bow shock in narrow-band H$\alpha$ imaging \citep{GvaBom08}, and the red supergiant IRC -10414 also has a bow shock detected in H$\alpha$ imaging \citep{GvaMenKni14}.
It seems likely that the bow shocks of Betelgeuse \citep{MohMacLan12} and $\zeta$ Ophiuchi \citep{GreMacKav22} could be detectable in nebular optical lines if the exceptionally bright stellar emission could be masked.

\begin{figure}
\begin{center}
 \includegraphics[width=1.0\textwidth]{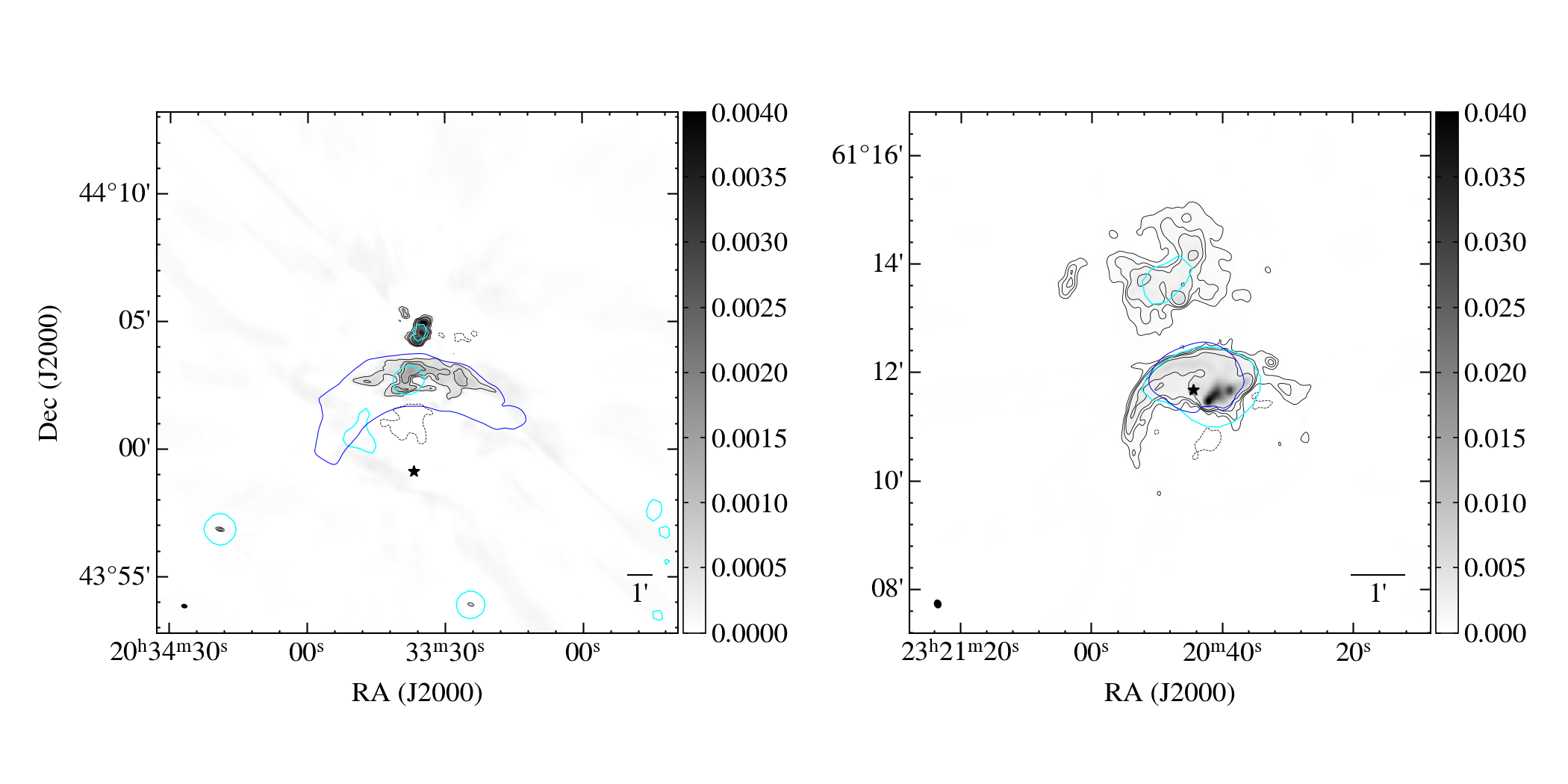} 
 \caption{Radio detection of the bow shocks around BD+43\,3654 (left) and BD+60\,2522 (right), obtained with the VLA at 4-8 GHz (greyscale and grey contours).  Blue contour shows \textit{WISE} band 4 emission in mid-IR, and cyan contours show 1.4\,GHz emission from the \textit{NVSS} survey.
   Credit: Moutzouri \textit{et al.}, 2022, Astronomy \& Astrophysics, 663, A80, fig.~1.}
   \label{fig:moutzouri22}
\end{center}
\end{figure}

The first radio detection of a bow shock was by \citet{BenRomMar10}, around the runaway O supergiant BD+43\,3654, further studied by \citet{BenDelHal21} and \citet{MouMacCar22} over a wider frequency range.
There is significant interest in radio studies at present because of the rapid advances in instrumentation and also the potential to detect both thermal (Bremsstrahlung) and non-thermal (synchrotron) radiation.
From its radio spectrum it seems that the bow shock of BD+43\,3654 exhibits both.
\citet{MouMacCar22} also detected the Bubble Nebula, NGC\,7635, as predicted by \citet{GreMacHaw19}, and the detections of this and BD+43\,3654 are shown in Fig.~\ref{fig:moutzouri22}.
Some low significance radio emission was found in NVSS survey data for E-BOSS bow shocks by \citet{PerBenIse15}, and some of these are now confirmed radio emitters with ASKAP observations \citep{VanSaiMoh22}.
The bow shock of Vela X-1 was also detected with MeerKAT \citep{VanHeyFen22}.
Given these recent discoveries and the promise of wide-field, broadband surveys with the new interferometric arrays, I anticipate many more radio detections of bow shocks in the coming years.

Thermal bremsstrahlung gives a direct measure of the emission measure, allowing to constrain the gas density in both the pre-shock and shocked ISM in the bow shock.
Obtaining gas density from IR emission has significant uncertainties arising from dust composition and grain size distribution \citep{PavKirWie13, MacHawGva16}, and nebular line emission can have uncertainties from patchy extinction along the line of sight.
Radio has the distinct advantage that is it unaffected by extinction.

With a detection of synchrotron radiation at radio wavelengths one can make model-dependent predictions for synchrotron and inverse-Compton emission at X-ray and $\gamma$-ray energies \citep{DelRom12, DelBosMul18, MouMacCar22}.
Searches for non-thermal X-rays \citep{ToaOskGon16, ToaOskIgn17} and 
$\gamma$-rays \citep{SchAckBue14, HESS2018_Bowshocks} from bow shocks have so far only obtained upper limits, although this is not altogether surprising given the predictions and sensititivies of current instruments.
Detection of high-energy radiation may need to wait for the next generation of instruments for bow shocks around single stars \citep{MouMacCar22}.

\section{The wind-ISM boundary layer -- theoretical models}

The contact discontinuity between shocked stellar wind and interstellar material is very strong for hot stars with fast winds.
Typical post-shock temperature of the shocked wind is $T_\mathrm{PS}\sim 10^8$\,K for an adiabatic shock, easily obtainted from the Rankine-Hugoniot jump conditions:
\begin{equation}
T_\mathrm{PS} \approx 5.6\times 10^7\,\mathrm{K} \left( \frac{v_\mathrm{sh}}{2000\,\mathrm{km\,s}^{-1}} \right)^2 \;,
\end{equation}
where an exact expression depends on the gas composition and ionization state.
Given a photoionized ISM temperature of $T_\mathrm{ISM}\approx8000$\,K, this implies a temperature jump of a factor of $10^4$ across the contact discontinuity, and an associated density jump of the same factor in the opposite sense.
For comparison this is similar to the density ratio of air and rock.
It is easy to see that mixing a small volume of interstellar gas across the contact discontinuity can have a large effect on the density and temperature (and hence radiative emissivity at different energies) of the wind bubble or bow shock.

Electronic thermal conduction \citep{CowMcK77} transports heat from the hot and low-density gas phase across a contact discontinuity into a boundary layer of cooler and denser gas.
The thickness of this boundary layer is determined by the mean free path of electrons, or the gyroradius in the presence of a magnetic field component parallel to the discontinuity.
The heated boundary layer expands, resulting in a region with strong temperature and density gradients rather than a well-defined contact discontinuity.
This can be seen in hydrodynamic bow shock simulations by \citet{ComKap98} and \citet{MeyMacLan14}, where the structure of the bow shock is signficantly modified by thermal conduction.
\citet{MeyMigKui17} showed, however, that the inclusion of an interstellar magnetic field dramatically reduces the effects of thermal conduction and, given that the ISM is pervaded by magnetic fields, one may deduce that the hydrodynamic models of wind bubbles \citep{WeaMcCCas77} and bow shocks \citep{ComKap98,MeyMacLan14} significantly overestimate the effects of thermal conduction.

\begin{figure}
\begin{center}
 \includegraphics[width=1.0\textwidth]{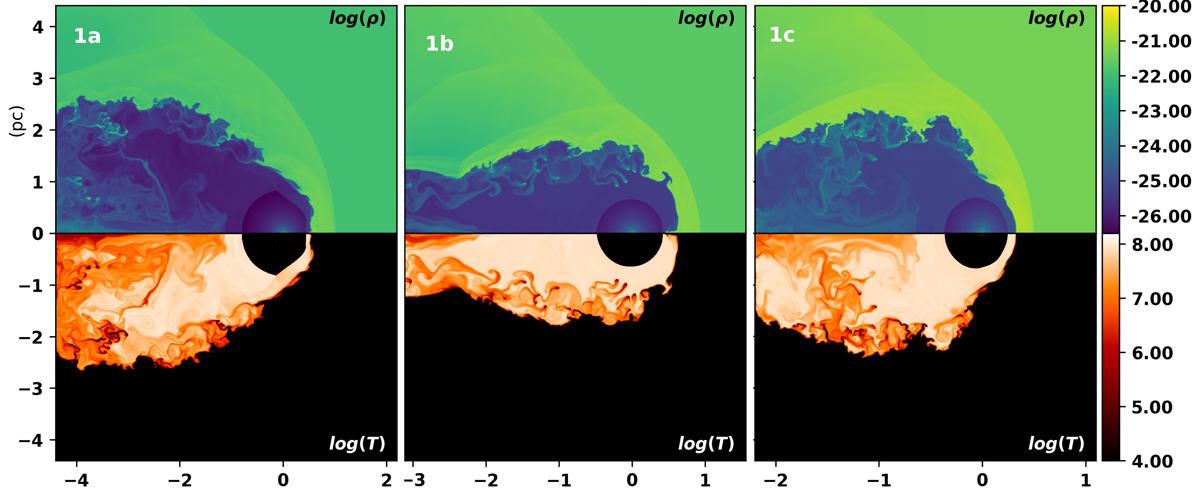} 
 \caption{Gas density (g\,cm$^{-3}$, upper half-plane) and temperature (K, lower half-plane) on a logarithmic colour scale for high-resolution 2D axisymmetric simulations of bow shock models for the Bubble Nebula, NGC\,7635.  Credit: Green \textit{et al.}, 2019, Astronomy \& Astrophysics, 625, A4, fig.~2.}
   \label{fig:green19}
\end{center}
\end{figure}

An alternative physical process that can thicken and distort the wind-ISM boundary layer is often referred to as dynamical or turbulent mixing \citep[e.g.][applied to supernova remnants]{SlaShuBeg93}.
This is generally driven by Kelvin-Helmholtz instability (KHI) because the hot phase has a sound speed up to $100\times$ that of the warm ISM phase, and so the velocity shear at the interface is typically large.
It cannot be captured in 1D simulations, but 2D simulations with an unstable wind-ISM interface (due to e.g.~Rayleigh-Taylor instability or non-linear thin-shell instability) or with an anisotropic external pressure naturally generate such shear flows \citep{ToaArt11, MacGvaMoh15, GreMacHaw19}.
This KHI-induced turbulent mixing is clearly shown in Fig.~\ref{fig:green19} (taken from \citealt{GreMacHaw19}), where waves initiated near the apex of the bow shock grow to non-linear amplitudes and the resulting vortices strongly mix wind and ISM material in the wake behind the bow shock.

If thermal conduction is not explicitly modelled then the degree of mixing is dependent on numerical resolution through both numerical diffusivity and the degree to which the numerical scheme has the resolution to capture the development of KHI \citep[e.g.][]{GreMacKav22}.
Recently, \citet{LanOstKim21a} and \citet{LanOstKim21b} developed a theory of turbulent mixing in wind bubbles expanding into clouds with a fractal density structure, arguing that this should lead to strongly cooled wind bubbles.
The degree of energy dissipation is a key constraint for larger scale simulations of the ISM of galaxies, for which it is currently not clear how efficient wind feedback (implemented in a sub-grid model) should be \citep[e.g.][]{FicGraRom22}.

\section{The wind-ISM boundary layer -- observations}

One way to constrain these processes observationally is to study the thermal X-ray or UV emission from the wind-ISM interface.
The brightest X-ray emitting wind bubbles are around Wolf-Rayet (WR) stars \citep{ToaGueChu15}.
These are wind-wind interactions where the fast wind of the WR star sweeps up the slow wind \citep{GarMacLan96} or mass ejection from binary systems \citep{JimArtToa21}.
Here there is usually considerable uncertainty about the density structure of the wind in the previous evolutionary phase.
UV lines from intermediate temperature gas at the hot-cold interface were also detected by \citet{BorMcCCla97} within a WR ring nebula and interpreted as a conduction front.

So far the only clear detection of diffuse X-rays from stellar wind of a single main-sequence star is very weak emission from within the bow shock around $\zeta$ Oph \citep{ToaOskGon16, GreMacKav22}.
Diffuse X-rays were also predicted \citep{MacGvaMoh15} and later detected \citep{TowBroGar18} from the massive star forming region RCW\,120, which likely includes a contribution from the O star that is the main driver of the nebula's evolution.
A quantitative comparison between theory and observation for this object has not been made.

On larger scales, X-ray emission has been detected from a number of massive star-forming regions \citep{TowBroGar18}.
The energy content of the hot phase was derived from observational data and has been compared with the energy input from winds by \citet{RosLopKru14}, finding that most of the input energy cannot be accounted for.
This is strong observational evidence for the effectiveness of energy transport across the contact discontinuity by e.g.~turbulent mixing, although `leakage` of energy into the low-density coronal gas surrounding star-forming regions also could not be excluded \citep{RogPit14}.

\section{3D MHD models of bow shocks and astrospheres}

We have seen that magnetic fields are important for moderating the effects of thermal conduction at the contact discontinuity of bow shocks, and it is also known that the operation of KHI is markedly different in the presence of a magnetic field \citep{FraJonRyu96, KepTotWes99}.
Axisymmetric simulations in 2D do not allow a general magnetic field configuration because only ISM fields that are parallel with (or anti-parallel to) the velocity vector of the star are permitted.
There is therefore a need for 3D MHD simulation of bow shocks, and a number of groups have worked on this over the past few years, developing efficient methods that enable 3D simulations with reasonable computational resources.

The computationally cheapest calculations are for stars with slow winds (because the timestep is inversely proportional to the fastest speed on the domain), and so the first 3D hydrodynamic simulations were for isothermal calculations of winds from cool stars \citep{BloKoe98}.
These showed very strong dynamical instability that quickly grew to nonlinear amplitudes.
More detailed calculations by \citet{MohMacLan12} applied to the red supergiant Betelgeuse explored the stability of the bow shock as a function of space velocity of the star, with a detailed radiative cooling prescription.
Recently \citet{MeyMigPet21} added magnetic fields to these models of bow shocks around cool stars, showing that the field has a strong stabilising influence on the bow shock.

Simulations of bow shocks around O stars are much more challenging because of the larger Mach number of the shocks (affecting stability of the scheme) and the timestep constraint (increasing the computational cost).
\citet{SchBaaFic20} developed a 3D MHD scheme using a spherical coordinate system that naturally has varying spatial resolution and hence is quite efficient.
They studied the differences between hydrodynamic and MHD simulations for a bow shock similar to that around the massive stars $\lambda$ Cep.
\citet{BaaSchKle21} used this model to study the effects of ISM inhomogeneities on the structure and observable properties of bow shocks.

\begin{figure}
\begin{center}
 \includegraphics[width=0.47\textwidth]{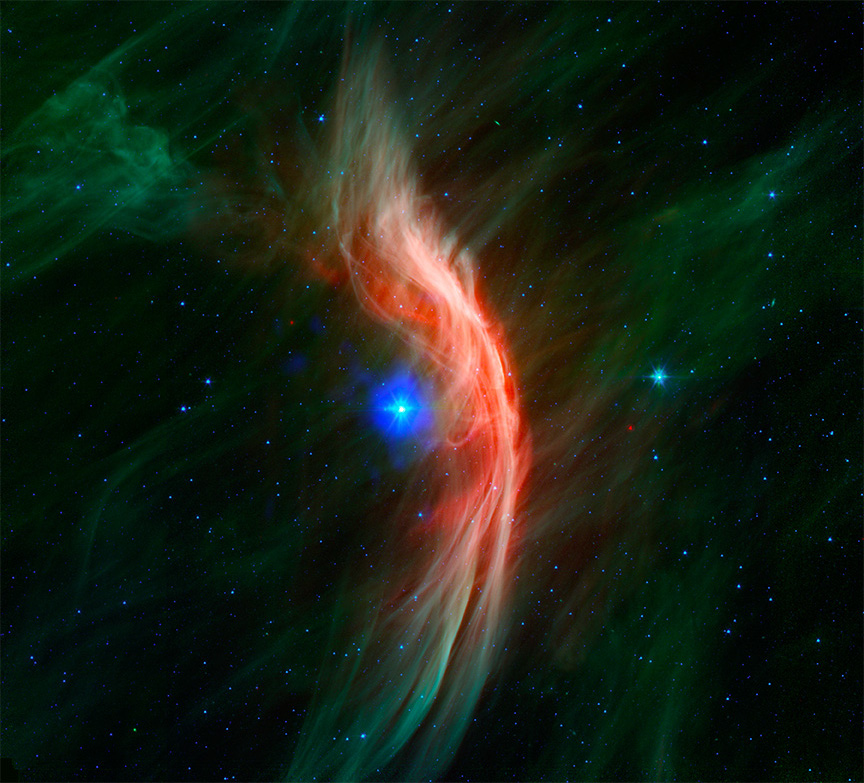} 
 \includegraphics[width=0.51\textwidth]{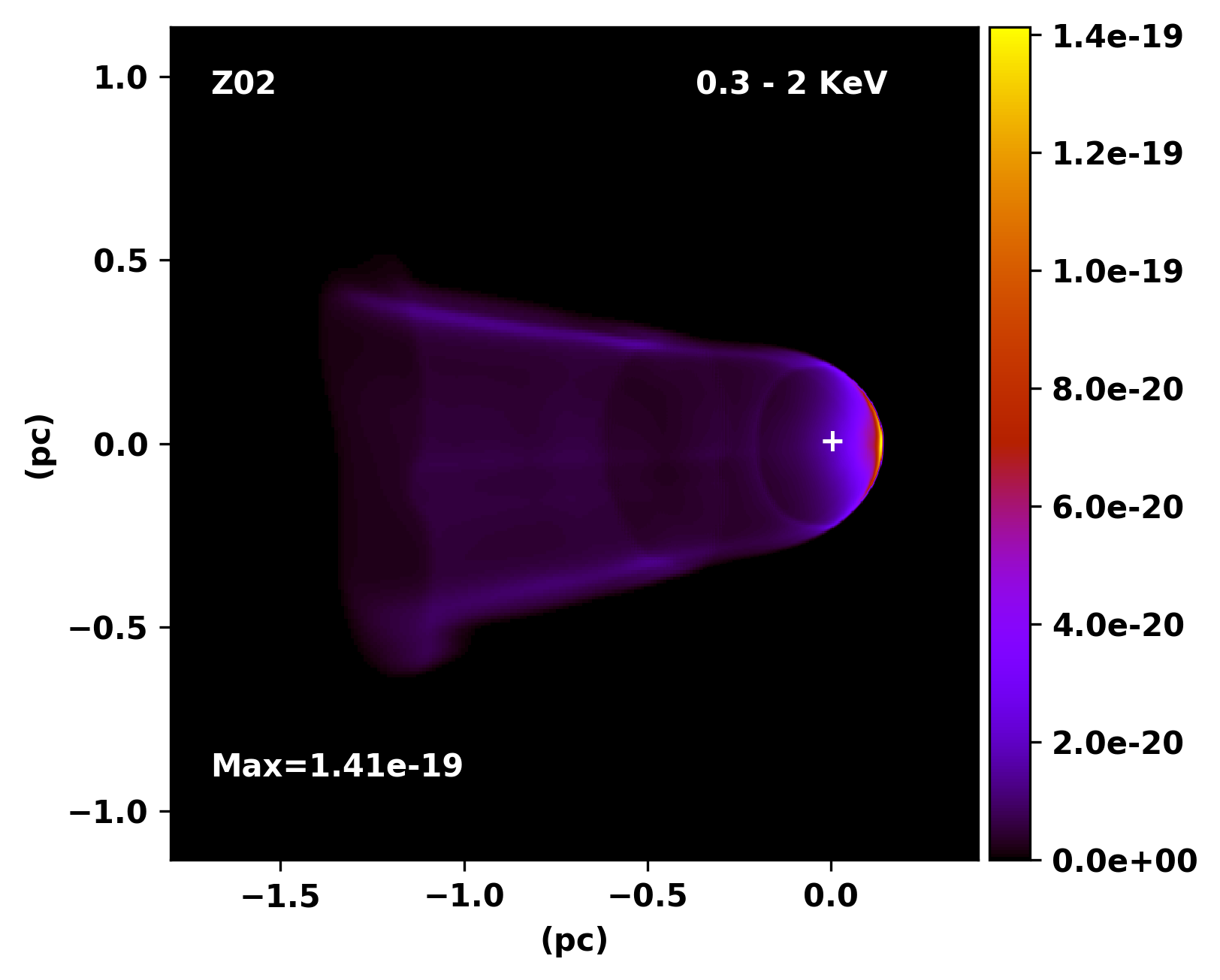} 
 \caption{\textbf{Left:} The bow shock of $\zeta$ Ophiuchi, seen in infrared (red and green, from \emph{Spitzer Space Telescope}) and X-rays (blue, \emph{Chandra X-ray Observatory}).  Image credit: X-ray: NASA/CXC/Dublin Inst.~for Advanced Studies/S.~Green et al.; Infrared: NASA/JPL/Spitzer.
  \textbf{Right:} synthetic X-ray emission from 3D MHD simulation of the bow shock, from 
  \citet{GreMacKav22} (Astronomy \& Astrophysics, 665, A35, fig.~13).}
   \label{fig:green22}
\end{center}
\end{figure}

Using a Cartesian coordinate system with static mesh-refinement, \citet{MacGreMou21} introduced 3D MHD simulations of bow shocks with the software \textsc{pion}.
This was used by \citet{GreMacKav22}, who presented the first 3D MHD simulations dedicated to modelling the bow shock of $\zeta$ Oph.
This spectacular system (see Fig.~\ref{fig:green22}) provides arguably the cleanest laboratory for testing mixing processes at the wind-ISM boundary. 
It is a very nearby massive star (136\,pc) with well-characterised proper motion, moving through a large photoionized H~\textsc{ii} region and driving a bright and well-resolved bow shock.
Crucially, diffuse X-ray emission has been detected with \emph{Chandra} \citep{ToaOskGon16}.
This source is also far out of the Galactic Plane, which can be valuable for multi-wavelength studies because background emission is less of an issue.
Becuase of rapid stellar rotation the radial velocity of the star is actually poorly constrained, and this introduced some uncertainty to the modelling.
Nevertheless the authors were able to show that the X-ray emission predicted by the simulated bow shock is somewhat weaker than the observed emission.
There are uncertainties, particularly the role of instabilities and effects of limited numerical resolution, but this comparison of 3D MHD simulations with high-resolution observations is a promising avenue for obtaining detailed information on the physical processes operating at shocks and contact discontinuities in the hot and warm ISM phases.

\section{Outlook}

The interaction of stellar winds with the ISM is important for interpreting observations of circumstellar nebulae, but also has broader application to a number of other areas of astrophysics.
The effectiveness of thermal conduction in astrophysical plasmas is so far only experimentally measured in the Solar wind \citep[e.g.][]{BalPulSal13}, and the degree to which it is inhibited by both large-scale coherent and small-scale turbulent magnetic fields can potentially be investigated at the wind-ISM interface.
It is important to constrain mixing of interstellar matter into stellar-wind bubbles and how this affects the efficiency of mechanical feedback from winds to the ISM, with consequences for models of galaxy formation and evolution.
The wind-ISM interaction sets the initial conditions into which a supernova blast wave expands, which can have important consequences for the evolution of supernova remnants \citep{DasBroMey22}.
Bow shocks and wind bubbles can be a useful laboratory for studying particle acceleration in shocks \citep{BenRomMar10, HESS2022_Westerlund1} and associated particle transport and radiation processes \citep{DelPoh18}.

This review has focussed on wind bubbles and bow shocks around single stars, but I want to emphasise that the coming decade will see great advances in modelling and observing wind-ISM and wind-wind interactions of binary and higher-order multiple star systems.
Unlike low-mass stars, almost all massive stars begin their lives in binaries and the majority will undergo interaction with a companion during their lifetime \citep{SanDeMdeK12}.
Colliding-wind binaries are one of the few astrophysical systems where time-dependent shock physics can be probed \citep{PitDou06, HESS20_EtaCar}, and they are bright Galactic sources across the electromagnetic spectrum up to TeV gamma rays.

Progress in our understanding of wind-ISM interaction depends on both new data and more detailed models.
In this regard we can anticipate breakthroughs within the next decade.
The rapid improvement in large field-of-view radio interferometry driven by \emph{SKA} pathfinder instruments is leading to detection of both thermal and non-thermal radio emission from circumstellar nebulae and bow shocks.
This gives important insights into gas density and thermal state, and the population of non-thermal particles.
In the next few years the \emph{Cherenkov Telescope Array (CTA)} will come online, providing unprecedented sensitivity to TeV gamma-ray emission, with potential detections of populations of very-high-energy particles accelerated in wind bubbles and bow shocks, and detection of a large population of Galactic binary systems, potentially including the colliding-wind binaries.
Looking further ahead, the \emph{ATHENA} mission promises huge sensitivity improvements compared with current X-ray telescopes, leading to much better characterisation of the hot thermal plasma of the shocked stellar-wind, and potentially detection of non-thermal emission via synchrotron or inverse-Compton radiation.

At the same time, rapid improvements in software for astrophysical fluid dynamics, including open-source community projects, mean that high-fidelity simulations of bow shocks and wind bubbles can now be performed.
In comparison with new high-resolution datasets coming from observations across the electromagnetic spectrum, these models have significant power to constrain uncertain physical parameters, discriminate between different physical models, and in general give a deeper and clearer understanding of astrophysical gas dynamics.

\section*{Acknowledgements}
JM acknowledges support from a Royal Society-Science Foundation Ireland University Research Fellowship and an Irish Research Council (IRC) Starting Laureate Award, and the DJEI/DES/SFI/HEA Irish Centre for High-End Computing (ICHEC) for computational facilities and support.
It is a pleasure to acknowledge the contributions of members of my research group at DIAS to this review through the papers cited and through many discussions over the past 7 years.
I am very grateful to the SOC of IAUS\,370 for the invitation to present this review at the IAUGA 2022 in South Korea.

\bibliographystyle{apj}
\bibliography{./refs}

\end{document}